\documentclass[twocolumn,showpacs,prb,superscriptaddress]{revtex4}
\usepackage{graphicx}
\usepackage{subfigure}

\def\oop{out-of-plane}
\def\ip{in-plane}
\def\ea{\emph{et al.}}

\begin{document}

\title{Influence of magnetic on ferroelectric ordering in LuMnO$_3$}

\author{Bas B. \surname{Van Aken}}
\email{bbv20@cam.ac.uk}
\affiliation{Department of Materials
Science and Metallurgy, University of Cambridge, UK}
\affiliation{Solid State Chemistry Laboratory, Materials Science
Centre, University of Groningen, The Netherlands}

\author{Thomas T. M. Palstra}
\affiliation{Solid State Chemistry Laboratory, Materials Science
Centre, University of Groningen, The Netherlands}

\date{\today{}}

\begin{abstract}
We have studied the influence of antiferromagnetic ordering on the
local dielectric moments of the MnO$\mathrm{_5}$ and LuO$_{7}$
polyhedra by measuring neutron powder diffraction patterns of
LuMnO$_3$ at temperatures near $T_N$. We show that the coupling is
weak, because the magnetic exchange coupling is predominantly in
the $ab$ plane of the MnO$_5$ trigonal bipyramids, and the
electric dipole moments, originating in the LuO$_7$ polyhedra, are
oriented along the hexagonal $c$ axis. Anomalies in the dielectric
properties near $T_N$ are thus caused by the geometric constraints
between the MnO$_5$ and the LuO$_7$ polyhedra.
\end{abstract}
\maketitle

Multiferroic magnetoelectrics are materials that combine
ferroelectric and magnetic ordering. These materials are of
fundamental and technological interest as coupling between
different order parameters enables multiple ways to interact with
these materials, such as pressure, electric and magnetic fields.
In most ionic materials the Coulomb repulsion between ions results
in centred (i.e. non-polar) crystal structures. Conventionally,
only for ions with lone pairs and non-magnetic ions, sizeable
dipole moments can exist\cite{Hil00}. For few extraordinary
materials magnetic and electric dipolar order can coexist. The
coupling between the magnetic and electric order must originate
either from magnetic ordering affecting the electric dipoles via
striction, or the displacements at the dipolar ordering changing
the magnetic order. These interactions depend intimately on the
coupling between the magnetic and electric building blocks of the
3D-crystal structure.

Recently the distorted perovskites BiMnO$_3$\cite{Mor02a,Ses03a},
BiFeO$_3$\cite{Wan03a} and also TbMnO$_3$\cite{Kim03c} have been
shown to combine magnetic and ferroelectric order. Another class
of materials that combine magnetic and electric dipolar order are
the hexagonal manganites with general formula AMnO$_3$, which are
antiferromagnetic and ferroelectric. Hexagonal AMnO$_3$
(h-AMnO$_3$), in particularly YMnO$_3$, has attracted interest for
applications, for instance in non-volatile memory
devices\cite{Fuj96} or as ferroelectric gate
electrode\cite{Ito03a}. Progress has also been made in the
understanding of the origin of the ferroelectric
properties\cite{Van04c} and the coupling between the ferroelectric
and antiferromagnetic order\cite{Fie02b,Han03a}. Fiebig \ea\ have
shown that below the antiferromagnetic ordering temperature $T_N$,
the ferroelectric domain walls, or twin  boundaries, always act as
antiferromagnetic (AFM) domain walls as well. Furthermore, AFM
domain walls also exist in the bulk of the ferroelectric (FE)
domains. This suggests  that not the orientation of the FE order
and the AFM order are linked, but that the transition from one FE
orientation to the other forces or accommodates a transition in
the AFM order\cite{Fie02b,Gol03a}. The coupling between electric
and magnetic ordering requires detailed knowledge of the crystal
structure. Unfortunately, there are conflicting  results in the
literature on the atomic positions when comparing neutron powder
diffraction (NPD) and single crystal X-ray diffraction (SXD).
Neutron diffraction results show anomalous variations in the bond
lengths\cite{Mun00,Kat02a} with temperature.

The crystal structure of h-AMnO$_3$ has two major differences
compared with the perovskite structure. Firstly, the Mn$^{3+}$
ions (high spin $3d^4$) are located at the centre of a trigonal
bipyramid instead of an octahedron. The resulting crystal field
splits the $3d$ levels in three energy levels, in order of
increasing energy: $xz$ and $yz$; $xy$ and $x^{2}-y^{2}$; and
$3z^{2}-r^{2}$. Consequently, Mn$^{3+}$ $3d^{4}$ has no partially
filled degenerate levels and is therefore \emph{not} Jahn-Teller
active\cite{Van01b}. Density of states calculations show some
mixing between the Mn $3d$ and O $2p$ levels.\cite{Van04c}
Secondly, the MnO$_5$ polyhedra are corner linked in sheets,
separated by a layer of A ions. This contrasts the MnO$_{6}$ 3D
network in perovskite manganites. As a result, h-AMnO$_3$ is
pseudo-layered, reflected for instance in the Mn spin being in the
$xy$ plane even above $T_N$\cite{Van01b,Lot01,Lot02a}.

To explain the how the AFM and FE order are linked, we studied the
temperature dependence of the crystal structure of LuMnO$_3$ around
the antiferromagnetic ordering temperature, $T_N=88$ K, which is much
lower than the FE ordering temperature, $T>573$ K\cite{Ber63a}.
Furthermore, we discuss the relation between the ionic radius of the A
ion $r_A$ and the crystal structure.

\section{Temperature dependence of L\lowercase{u}M\lowercase{n}O$_3$}
LuMnO$_3$ has been prepared with standard solid state synthesis.
Neutron powder diffraction has been performed at the Polaris
time-of-flight instrument at ISIS. The data have been refined using
the \textsc{gsas} package\cite{GSAS}, including isotropic temperature
factors. Sample quality has been checked by measuring a neutron powder
diffraction pattern at ambient temperature. No reflections from
impurity phases could be detected and the refined structure was in
good agreement with single crystal data\cite{Van01e}. Temperature
dependent measurements, between 40 and 120 K, were performed at the
same beam line, using a standard Orange cryostat. Typically quality
factors for the refinements are $\chi^{2}=2.963$, R(F$^{2})=4.79\%$
and R$_{\textrm{p}}=1.74\%$ for 47 variables.

\begin{figure}[htb]
\subfigure[]{\includegraphics[bb=30 670 150 820,
width=60mm]{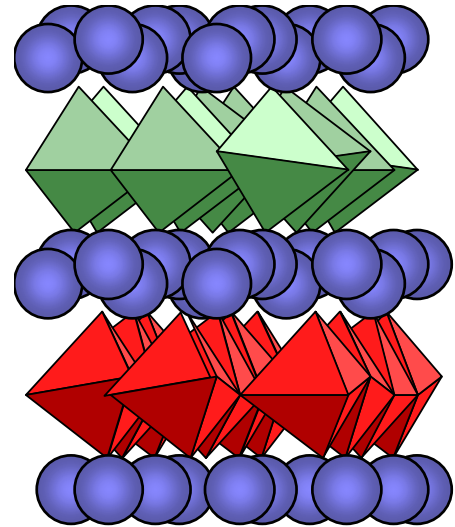}\label{Fig:Layers}}
\subfigure[]{\includegraphics[bb=45 210 560 645,
width=75mm]{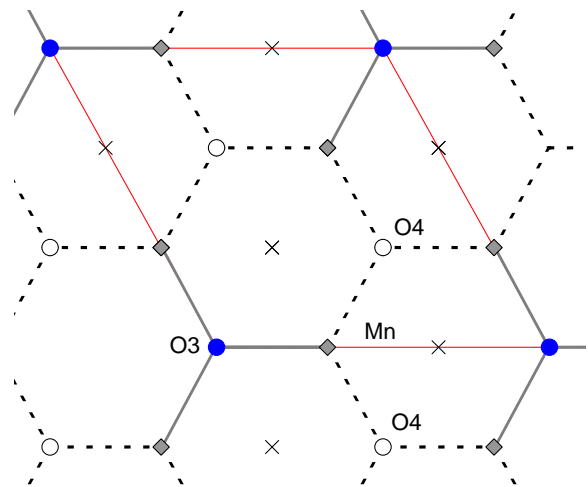}\label{Fig:MnObonds}}
\caption{\subref{Fig:Layers} View perpendicular to the $c$ axis of
the crystal structure of AMnO$_3$. The trigonal bipyramids
represent the MnO$_5$ and the circles the A ions. This sketch
highlights the stacking of the MnO$_5$ layers.
\subref{Fig:MnObonds} Sketch of the z=0 layer of LuMnO$_3$,
showing the two inequivalent Mn - O bonds. Mn, O3 and O4 are shown
as grey diamonds, closed and open circles, respectively. Mn - O3
and Mn - O4 bonds are indicated by solid and dashed lines,
respectively. The thin line outlines the unit cell.}
\end{figure}

Fig.~\ref{Fig:Layers} shows the crystal structure of hexagonal
manganite. It consists of non-connected layers of MnO$_5$ trigonal
bipyramids, that are corner-linked by \ip\ oxygen ions, O3 and O4.
The apical oxygen ions, O1 and O2, form two close packed layers
separated by a distorted layer of A$^{3+}$ ions. In
Fig.~\ref{Fig:MnObonds} a cross-section through the $z\sim0$ plane
is given. The oxygen positions O1 and O2 are similar to the Mn
positions, albeit with $z\sim\pm\frac{1}{6}$.  The A site ions A1
(A2) are located exactly above the positions O3 (O4). In the next
Mn ions containing layer ($z\sim\frac{1}{2}$), the Mn, O1 and O2
positions are located at \textsf{x} (Fig.~\ref{Fig:MnObonds}).

The spacegroup of hexagonal LuMnO$_3$ in the ferroelectric phase is
$P6_3cm$. The hexagonal unit cell contains six formula units; there
are seven inequivalent positions with 10 refinable parameters. Since
all symmetry elements are parallel to the $c$ axis, all $z$
coordinates are ``free''. For practical reasons, we have chosen to fix
$z_{Mn}=0$. The atomic positions are listed in Table~\ref{Tab:atompos}.

\begin{table}[htb]
\caption{The atomic positions of hexagonal manganite AMnO$_3$.
\emph{m} is the multiplicity of the site.} \label{Tab:atompos}
\begin{ruledtabular}
\begin{tabular}{c|rrrc}
 atom&\emph{x} (-)& \emph{y} (-)& \emph{z} (-)& \emph{m}\\ A1& 0& 0&
 $z_{A1}\sim\frac{1}{4}$& 2 \\ A2& $\frac{1}{3}$& $\frac{2}{3}$&
 $z_{A2}\sim\frac{1}{4}$& 4\\ Mn& $x_{Mn}\sim\frac{1}{3}$& 0&
 $z_{Mn}=0$& 6\\ O1& $x_{O1}\sim\frac{1}{3}$& 0&
 $z_{O1}\sim\frac{1}{6}$& 6\\ O2& $x_{O2}\sim\frac{1}{3}$& 0&
 $z_{O2}\sim-\frac{1}{6}$& 6\\ O3& 0& 0& $z_{O3}\sim0$& 2\\ O4&
 $\frac{1}{3}$& $\frac{2}{3}$& $z_{O4}\sim0$& 4\\
\end{tabular}
\end{ruledtabular}
\end{table}

\begin{figure}[htb]
\includegraphics[bb=39 425 273
600,width=75mm]{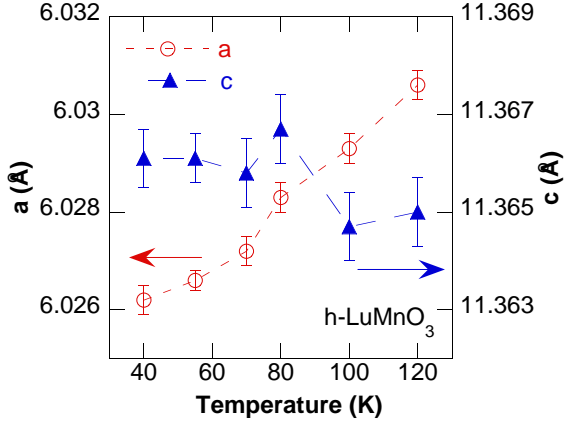} \caption{Temperature dependence of lattice
parameters of LuMnO$_3$.\label{Fig:LatPar}}
\end{figure}

In Fig.~\ref{Fig:LatPar} the lattice parameters has been plotted
versus the temperature. The $a$ axis decreases with decreasing
temperature at a rate of $8\cdot10^{-6}$ K$^{-1}$. The lattice
parameter $c$ is nearly independent of $T$, although a small
discontinuity in $c$ ($\sim0.001$ \AA) can be seen near $T_N$. SXD
data at 295 K\cite{Van01e}, $a=6.038(1)$ \AA, are in agreement
with the extrapolation of Fig.~\ref{Fig:LatPar}. Thermal expansion
data have been reported on powdered YMnO$_3$ samples at higher
temperatures\cite{Les01a}. The expansion coefficient $\sim
7\cdot10^{-7}$ K$^{-1}$ obtained from these data is one order of
magnitude smaller than the value we find.

\begin{figure}[htb]
\includegraphics[bb=39 480 263
650,width=75mm]{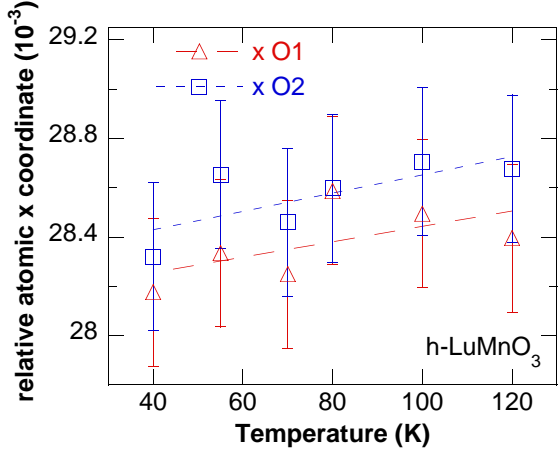} \caption{Temperature dependence of the
$x$ coordinate of the apical oxygen positions of LuMnO$_3$. The
absolute difference with the paraelectric position at
$x=\frac{1}{3}$ is plotted. Triangles and squares depict O1 and
O2, respectively. The dashed lines are least squares linear fits.}
\label{Fig:Oippos2}
\end{figure}

Compared to the paraelectric centrosymmetric structure, the
MnO$_5$ bipyramids are rotated along an axis through the Mn ion
and parallel to the O4-O4 bond. This buckling moves the O3 (O4)
ions up (down) along the $z$-axis and moves the O1(O2) ions at
($\sim \frac{1}{3}$, 0, $\pm \frac{1}{6}$) along the $x$-axis.
Fig.~\ref{Fig:Oippos2} shows the $T$ dependence of the apical
oxygen positions O1 and O2 of LuMnO$_3$, plotted as the absolute
difference with their paraelectric position at $x=\frac{1}{3}$.
The $x$ coordinate of these oxygen ions is a measure for the
buckling of the MnO$_5$ trigonal bipyramid, but small changes in
$x_{O1}$ and $x_{O2}$ do not have a significant effect on the
Mn-O1 and Mn-O2 bond lengths. The linear fits suggest that the
buckling might reduce with decreasing $T$, but the slope and its
error are the same order of magnitude. The angle of the line
through O1 and O2 with the $xy$-plane changes from $84.7^\circ$ at
$T=120$ K to $84.8^\circ$ at $T=40$ K. We conclude that there is
hardly any change in the buckling of the MnO$_5$ layers with
temperature.

The \oop\ A1$-$O3 and A2$-$O4 bond lengths are calculated using:
\begin{equation}
\textrm{A1$-$O3}=c\hskip2pt(z_{A1}-z_{O3})
\label{Eq:A1O3}
\end{equation}
and
\begin{equation} \textrm{A2$-$O4}=c\hskip2pt(z_{A2}-z_{O4}).
\label{Eq:A2O4}
\end{equation}
Since the temperature dependence of lattice parameter $c$ is
negligible, the $T$ dependence of these bond lengths depend only on
the $T$ dependence of the $z$ coordinates. We found in our refinements
that the $T$ dependencies of these $z$ coordinates have strong
positive correlations. Any changes in any of the $z$ coordinates is
mimicked closely by its companion. Therefore, the Lu1$-$O3 and
Lu2$-$O4 bond lengths show no dependence on temperature.

\begin{center}
\begin{figure}[htb]
\includegraphics[bb=30 490 310
779,width=75mm]{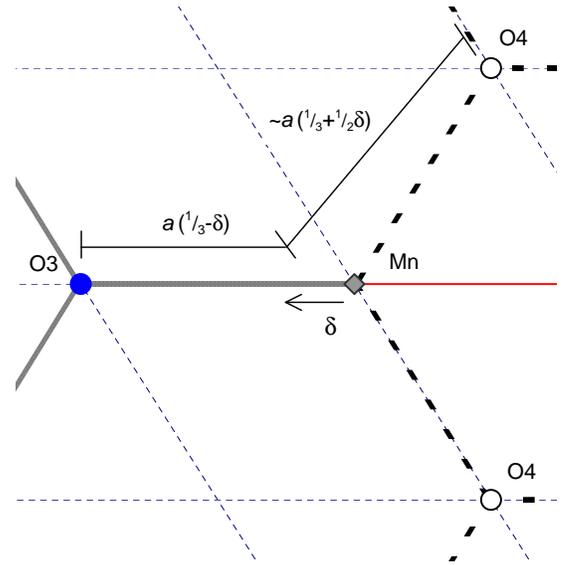} \caption{Sketch of the
relation of the length of the two inequivalent \ip\ Mn$-$O bonds,
showing the effect of an \ip\ Mn displacement. Symbols are the same as
in Fig.~\ref{Fig:MnObonds}. Thin dashed lines are lines parallel to
the unit cell axes at $\frac{1}{3}a$ intervals to guide the eye.}
\label{Fig:Mndisplacement}
\end{figure}
\end{center}

Each Mn ion has two Mn$-$O4 bonds and one Mn$-$O3 bond as is shown in
Fig.~\ref{Fig:MnObonds}. The \ip\ Mn$-$O3 bond length is given by
\begin{equation}
\textrm{Mn$-$O3}=\sqrt{a^{2}x_{Mn}^{2}+c^{2}z_{O3}^{2}}\sim
a\hskip2ptx_{Mn},
\label{Eq:MnO3}
\end{equation}
where in the approximation the effect of $z_{O3}\neq0$ can be ignored,
since the partial derivatives are proportional to the magnitude of
$x_{Mn}$ and $z_{O3}$. The other \ip\ bond length is given by
\begin{equation}
\textrm{Mn$-$O4}=\sqrt{a^{2}\{(x_{Mn}-1/2)^{2}
+(\sqrt3/6)^{2}\}+c^{2}z_{O4}^{2}}.
\end{equation}
Mn$-$O3 and Mn$-$O4 depend both on lattice parameter $a$ and on the
$x$ coordinate of the Mn ion. We can write
$x_{Mn}=\frac{1}{3}-\delta$, where $\delta$ is the displacement of the
Mn ion along the $x$ axis from its paraelectric position at
$(\frac{1}{3},0,0)$, see Fig.~\ref{Fig:Mndisplacement}.

Then, applying $x_{Mn}=\frac{1}{3}-\delta$, Mn$-$O3 equals
$a(\frac{1}{3}-\delta)$ and Mn$-$O4 equals
\begin{eqnarray}
\textrm{Mn$-$O4}=\sqrt{a^{2}\{(1/6+\delta)^{2}+1/12\}
+c^{2}z_{O4}^{2}}\sim \nonumber\\ \sim a\hskip2pt(1/3+1/2\delta),
\label{Eq:MnO4}
\end{eqnarray}
where the effect of $z_{O4}\neq0$ and $\delta^{2}$ is negligible.

\begin{figure}[htb]
\includegraphics[bb=45 480 263
659,width=75mm]{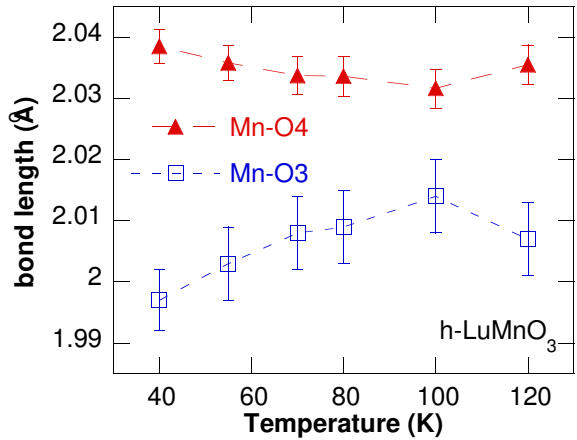} \caption{Temperature
dependence of \ip\ Mn$-$O bond lengths around $T_N=88$ K.}
\label{Fig:MnOip}
\end{figure}

From these approximate formulae we can see that a change, of $-\delta
a$, in the length of the Mn$-$O3 bond length, due to a displacement
$\delta$ of the Mn ion, is always accompanied with a change in the
length of the Mn$-$O4 bond of opposite sign and half the magnitude
($+\frac{1}{2}\delta a$). These \ip\ Mn$-$O bond lengths are plotted
against temperature in Fig.~\ref{Fig:MnOip}. Local extrema near $T_N$
are observed with the maximum value for Mn$-$O3 smaller than the
minimum for Mn$-$O4. Below and above $T_N$ the difference increases
between the \ip\ bond lengths. However the observed differences are
not  large, at most $0.041$ \AA.

The \oop\ Mn$-$O bond lengths are given by
\begin{equation}
\textrm{Mn-Oi}=\sqrt{a^{2}(x_{Mn}-x_{Oi})^{2}+c^{2}z_{Oi}^{2}}\sim
c\hskip2ptz_{Oi}; (i=1,2)
\end{equation}

where the $x$ components are neglected in the approximation, since
$x_{Mn}-x_{Oi}\ll z_{Oi}; (i=1,2)$. The \oop\ bond lengths also show
local extrema near $T_N$, but the changes are less pronounced than
for the \ip\ bond lengths. The difference between the inequivalent
\oop\ Mn$-$O bond lengths is smaller than $0.02$ \AA\ for all $T$.

The \oop\ bond lengths are much shorter than the \ip\ bond
lengths, respectively $\sim1.87$ \AA\ and $\sim2.02$ \AA. This is
related to the crystal field splitting of the Mn $3d$ orbitals,
with the unoccupied $3z^{2}-r^{2}$ orbital in the \oop\ direction.
The antibonding character of the Mn $3d$ - O $2p$ interaction
elongates the \ip\ bond lengths\cite{Van01b}.

We note that with SXD\cite{Van01a,Van01e,Van01f,Van01g}, it is found
that the Mn$-$O4 bond length is smaller than the Mn$-$O3 bond length,
whereas here we find that Mn$-$O4 is larger than Mn$-$03. However,
the difference between the inequivalent \ip\ bond lengths is small
($<0.05$~\AA). The \oop\ Mn$-$O bond lengths are in line with the SXD
data, and the difference between the inequivalent \oop\ bond lengths
is small ($<0.02$~\AA) in agreement with the SXD data. Other NPD
experiments on the crystal structure of hexagonal manganites in the
literature report larger differences between the \ip\ or the \oop\
Mn$-$O bond lengths. For instance, Katsufuji \ea\ report on LuMnO$_3$
at 300 K and find \oop\ bond lengths of 1.98~\AA\ and
1.78~\AA\cite{Kat02a}. Mu\~{n}oz \ea\ find for YMnO$_3$ at room
temperature a difference between \ip\ bond lengths
$\Delta_{\mathrm{in-plane}}$ of 0.1~\AA\cite{Mun00}. Also older
structure reports using SXD show large differences between \ip\
Mn$-$O bond lengths or \oop\ Mn$-$O bond lengths. Yakel reports
$\Delta_{\mathrm{in-plane}}=0.08$~\AA\ and
$\Delta_{\mathrm{out-of-plane}}=0.09$~\AA\ for LuMnO$_3$\cite{Yak63}.
Isobe reports $\Delta_{\mathrm{in-plane}}=0.048$~\AA\ and
$\Delta_{\mathrm{out-of-plane}}=0.058$~\AA\ for YbMnO$_3$ at $T=295$
K\cite{Iso91}. These SXD experiments however did not include
reflections from all regions of $hkl$ space, which is necessary for
SXD on non-centrosymmetric crystals.

In the previous section, we have discussed the temperature dependence
of the bond lengths. Now, we will investigate whether the observed
changes have any effect on the local dipole moments and the
macroscopic ferroelectric moment. In the remainder of this article,
we will define the local dipole moment associated with a cation site
as the distance between the centre of gravity of the nearest
neighbour O ions and the cation site.

In Fig.~\ref{Fig:MnOip} we have shown that the \ip\ Mn$-$O bond
lengths show local extrema near $T_N$, with the difference between
them smallest at $T_N$. This means that also their contribution to
the local dipole moments on the Mn site is smallest at $T_N$. The
space group symmetry relates the local dipole moment created by the
small \ip\ displacement of the Mn ion to the local dipole moments
originating from the other two Mn ions on the same plane in the unit
cell. Due to the sixfold screw axis, the local dipole moments are
rotated over an angle of 120$^{\circ}$ with respect to each other. As
a result, there cannot be any contribution to the macroscopic
ferroelectric moment by the \ip\ displacements.

The difference between the \oop\ Mn$-$O1 and Mn$-$O2 bond lengths
create a small local dipole moment $<0.02$~\AA\ from the Mn
coordination. These local dipole moments are parallel to the $c$
axis and they are not cancelled by the crystal symmetry. However
larger local dipole moments originate from the Lu1 and Lu2
coordination\cite{Van04c}. The local dipole moments from the Lu
coordination are at least an order of magnitude larger than the
ones from the Mn coordination.

\begin{figure}[htb]
\includegraphics[bb=40 420 275
595,width=80mm]{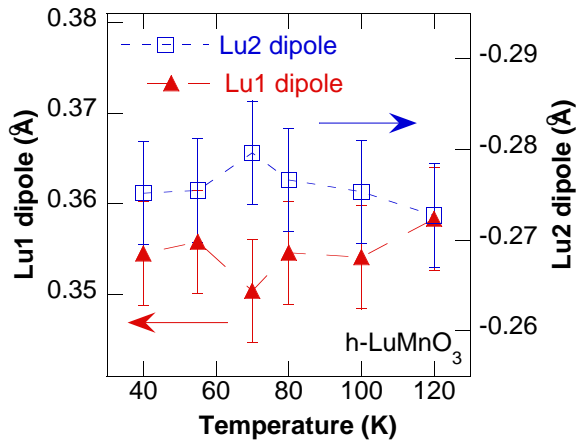}\caption{Temperature
dependence of LuMnO$_3$ of the local dipole moments from the Lu1 and
the Lu2 polyhedra near $T_N=$88 K. Note that the right axis is
reversed. \label{Fig:LuOdipole}}
\end{figure}

The local dipole moments associated with the LuO$_{8}$ polyhedra are
shown in Fig.~\ref{Fig:LuOdipole}. The calculation of the centre of
gravity of the oxygen polyhedra includes the seven ions at
$\sim2.4$~\AA\ and the one ion at $\sim3.2$~\AA. Note that the local
dipole moments from the Lu2 coordination are antiparallel with the
ones from the Lu1 coordination.

From Fig.~\ref{Fig:LuOdipole} it is clear that cooling through the
N\'{e}el transition has no significant influence on the local
dipole moments of LuMnO$_3$. However, we have shown in
Fig.~\ref{Fig:MnOip} that the local environment around the Mn ions
does change. Since the Mn spins are \ip\cite{Van01b}, even above
$T_N$, the magnetic ordering transition at $T_N$ is expected to
have the largest effect on the \ip\ Mn$-$O bond lengths. The
ferroelectricity on the other hand depends mostly on the \oop\
Lu$-$O bond lengths\cite{Van04c}. Although the oxygen ions
involved are the same, O3 and O4, the Mn$-$O bond lengths change
mostly by a displacement $\delta$ of the Mn ion along the $x$
axis, whereas the \oop\ Lu$-$O bond lengths change because of an
opposite movement of the Lu1 (Lu2) and O3(O4) ions along the $c$
axis. We conclude that there is no significant coupling between
the ferroelectric moment of LuMnO$_3$ and the antiferromagnetic
ordering temperature. This is in agreement with the conclusions
from Ref.~\onlinecite{Fie02b}.

\section{rare earth ionic radius}
In the second part of this article we will discuss the influence
of the rare earth ions, by their ionic radius $r_{A}$, on the
local dipole moments. The data used in this analysis has been
taken from Refs.~[\onlinecite{Van01a}] (YMnO$_3$);
[\onlinecite{Van01f}] (ErMnO$_3$); [\onlinecite{Van01g}]
(YbMnO$_3$) and [\onlinecite{Van01e}] (LuMnO$_3$).

\begin{figure}[htb]
\includegraphics[bb=40 420 275 595, width=80mm]{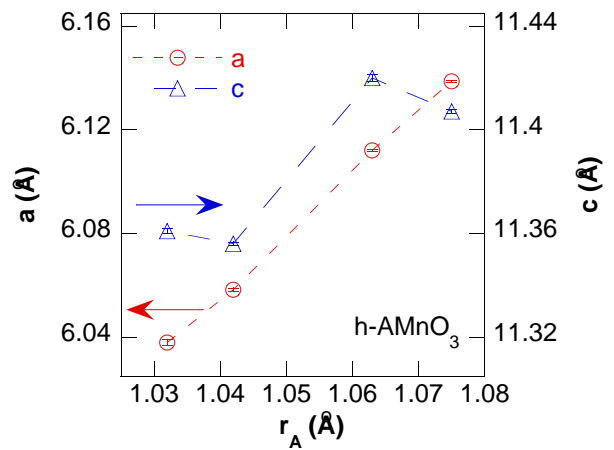}
\caption{Lattice parameters for hexagonal AMnO$_3$ as a function
of $r_{A}$ (A=Lu, Er, Yb and Y) at $T=295$ K.}\label{Fig:Rac}
\end{figure}

In Fig.~\ref{Fig:Rac}, the lattice parameters $a$ and $c$ at $T=295$ K
are plotted
against $r_A$. The values of $r_A$ have been taken from Shannon and
Prewitt \cite{Sha76}. We observe that with increasing $r_A$ the
lattice parameters increase. From LuMnO$_3$ to YMnO$_3$ lattice
parameter $a$ increases linearly, with a total increase of $\sim
1.7\%$. Lattice parameter $c$ has a smaller increase of about 0.5\%.
We can clearly see from these data that the influence of $r_A$ is
larger on the \ip\ lattice parameter than on the \oop\ lattice
parameter.

The crystal structure can be regarded as a stacking of hexagonally
packed oxygen layers and A ion layers. One expects that O$-$O bond
lengths are $\sim2.8$ \AA, ($r_O=1.4$ \AA) and A$-$O bond lengths
are $\sim2.4$ \AA, ($r_A\sim 1.05$ \AA). For instance, in
LaMnO$_3$ the average O$-$O and La$-$O bond length (in the
fcc-like closed packed OO and AO layers) is $\sim2.7$~\AA\
($r_{La}=1.2$~\AA). The observed values for the inter-plane A$-$O1
and A$-$O2 bond lengths in hexagonal AMnO$_3$ (not shown here) are
consistent with the expected values. Likewise the observed
increase with increasing $r_A$ of the A$-$O1 and A$-$O2 bond
lengths (on average 1.6\%) is in excellent agreement with the
increase of $a$ with $r_A$.

We note that in hexagonal AMnO$_3$ both the hexagonally packed O
layers at $z\approx\pm\frac{1}{6}$ and the MnO layers at
$z\approx0$ have intra-plane O$-$O distances of $\sim3.6$~\AA.
Similarly, the \ip\ A$-$A distances are also $\sim3.6$~\AA. This
is in contrast with the intra-plane bond lengths in pseudo-cubic
perovskites. These anomalously large "bond lengths" can be
understood, by considering the way the layers are stacked.
Starting at the MnO layer at $z=0$, the stacking order is MnO - O
- A - O - MnO with layer spacings of about $\frac{2}{12}c$,
$\frac{1}{12}c$, $\frac{1}{12}c$ and $\frac{2}{12}c$. One can
visualise that the A ion layer (at $z\approx\frac{1}{4}$) is
clamped between the closed packed O layer (at
$z\approx\frac{1}{6}$ and $z\approx\frac{1}{3}$). This will
enlarge the spacing of the O and A ions in those layers. The
larger the radius of the A ions, the more the O ions are pushed
apart. However, the O$-$O bond length between the two neighbouring
O layers at $z\approx\frac{1}{6}$ and $z\approx\frac{1}{3}$ is
reduced to $\sim2.8$~\AA, which corresponds to the value one
expects ($2\times r_O$). In this simple model, changes in $r_A$
can be accommodated completely by a change in the \ip\ bond
lengths and lattice parameter $a$.

\begin{figure}[htb]
\includegraphics[bb=45 420 265 595,width=80mm]{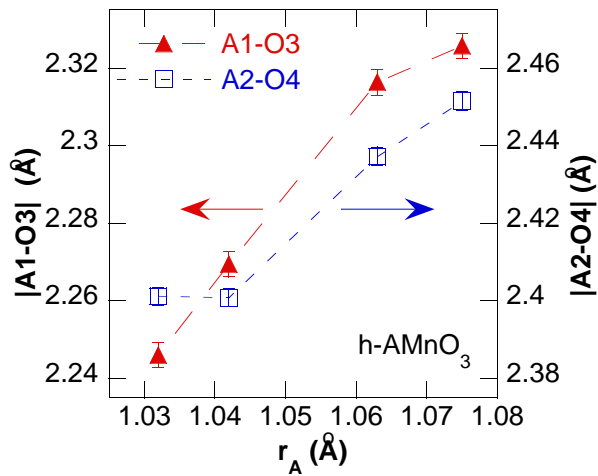}
\caption{\oop\ A$-$O bond lengths as a function of $r_A$.}
\label{Fig:RO}
\end{figure}

The decreasing distances in the $ab$ plane with decreasing $r_{A}$
should also be reflected in the \ip\ Mn$-$O bond lengths. The
change of $\sim1.6$ \% is in good agreement with the observed
change of $a$ with decreasing $r_{A}$. Possible changes in the
\oop\ Mn-O bonds are too small to be resolved within the scatter
and error bars of the experiments. However, we observe a distinct
relation between the A1$-$O3 and A2$-$O4 bond lengths and $r_A$,
as shown in Fig.~\ref{Fig:RO}.

\begin{figure}[htb]
\includegraphics[bb=45 420 265 595,width=80mm]
{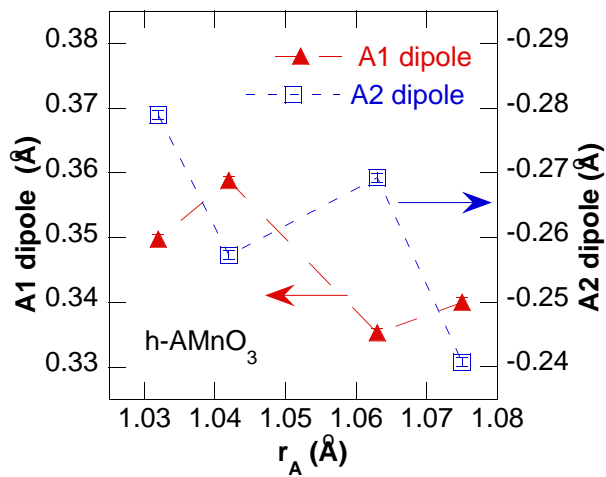} \caption{Dipole moments of the A1 and A2 polyhedra
of h-AMnO$_3$. Note that the A2 dipole is plotted on a reverse
scale. Both dipoles tend to increase with decreasing $r_{A}$.}
\label{Fig:R12dipole}
\end{figure}

The short A1$-$O3 and A2$-$O4 bonds increase by 3.6\% and 2.3\%
with increasing $r_A$. Therefore, the "extra long" \oop\ A$-$O
bond lengths should decrease with increasing $r_A$, since $c$ is
almost constant. Consequently, the dipoles of the A polyhedra
decrease with increasing $r_A$. The increase in dipole moment from
Y ($r_A=1.075$~\AA) to Lu ($r_A=1.032$~\AA) is about 10\% as shown
in Fig.~\ref{Fig:R12dipole}.

\begin{figure}[htb]
\includegraphics[bb=45 420 265 595,width=80mm]
{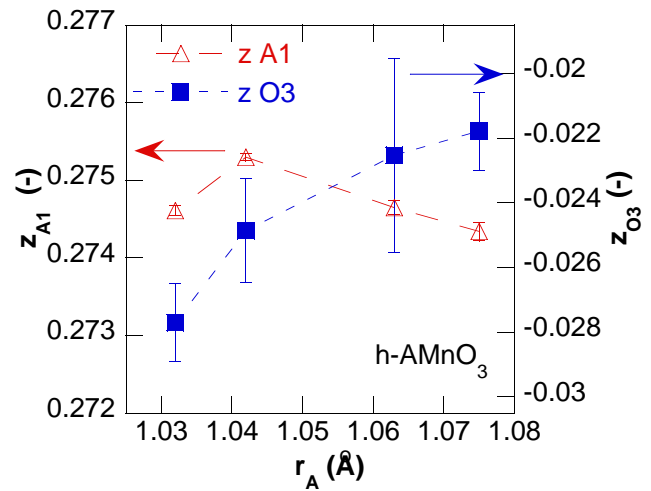} \caption{Fractional $z$ coordinates
of the A1 and O3 position (0, 0, $z$) versus $r_A$. Clearly the
$z$ coordinates deviate more from their equilibrium positions,
respectively $z=\frac{1}{4}$ and $z=0$, with decreasing $r_A$.}
\label{Fig:R1O3pos}
\end{figure}

The transition from the paraelectric centrosymmetric to the
ferroelectric non-centrosymmetric structure consists of two atomic
displacements. The first is the buckling of the MnO$_5$ network,
displacing O1 and O2 in the $xy$ plane and O3 and O4 along the $z$
axis. The second is the alternate displacement of the A1 and A2
ions along the $z$ axis away from the $z=\frac{1}{4}$~mirror
plane\cite{Van04c}. Formally, it has been shown that these atomic
displacements consist of two modes. Firstly, there is a K$_{3}$
mode, which changes the symmetry but can inherently not create a
macroscopic ferroelectric moment. Secondly, there is a $\Gamma$
mode, which does not change the space group symmetry, but
introduces the macroscopic ferroelectricity\cite{Lon03b,Lon03c}.
To see which of these two modes is influenced more by chemical
substitution, we have plotted in Fig.~\ref{Fig:R1O3pos} the $z$
coordinates of the A1 and the O3 sites. Both $z$ coordinates of A1
and O3 are about 0.02$c$ away from their centrosymmetric
equilibrium position. Clearly, the changes in the $z$ coordinate
of O3 site are much larger than the changes in the $z$ coordinate
of the A1 site. The data in Fig.~\ref{Fig:R1O3pos} therefore show
that upon chemical substitution of the A site the buckling
increases with decreasing $r_A$, which results in an increase of
the deviation from the centrosymmetric positions. The displacement
of the A ions from the mirror plane is more or less constant. This
is supported by the changes in the $x$ coordinates of the O1 and
O2 sites. Both are about 0.025$a$ off their symmetric position at
$x=\frac{1}{3}$ and this deviation becomes larger by $\sim 10\%$
from R=Y to R=Lu. We conclude that the change in the local dipole
moment is caused by atomic rearrangement and not by expansion of
the unit cell.

The ferroelectric transition temperatures for these materials have
been reported, but different authors report completely different
values or only "higher than" values. Furthermore, there are
indications that the ferroelectric ordering temperature and the
non-centrosymmetric to centrosymmetric phase transition temperature
are different by several hundreds degrees\cite{Luk74,Lon03b}. It is
therefore not possible to relate the magnitude of the local dipole
moments or the total dipole moment to the ferroelectric ordering
temperature.

The effect of changing $r_A$ can be summarised as follows. A
decrease in $r_A$ will reduce the ``normal'' A$-$O bond lengths.
This will change the \ip\ Mn$-$O bond lengths too. These changes
are accompanied by an increased buckling of the MnO$_5$
bipyramids, parameterised by $x_{O1}$, $x_{O2}$, $z_{O3}$ and
$z_{O4}$. The changes in the oxygen $z$ coordinates induce an
increase in the difference between the short and long A1$-$O3 and
A2$-$O4 bond lengths. This increases the dipole moments from the
A1 and A2 coordinations with decreasing $r_{A}$. However, the
antiparallel coupling between the local dipole moments at the A1
and the A2 site prevents a clear effect on the total dipole
moment.

In conclusion, the hexagonal AMnO$_3$ compounds exhibit both
antiferromagnetic and ferroelectric order. The magnetic exchange and
spins are confined in the basal plane, whereas the dielectric moments
originate predominantly from displacements parallel to the
$c$-axis. Consequently, the distortions due to magnetostrictive
effects and dipolar ordering are only weakly coupled.

We thank Ron Smith, Anne Dros and Auke Meetsma for experimental
assistance. This work is supported by the `Stichting voor
Fundamenteel Onderzoek der Materie (FOM)', NWO for neutron beam time
at ISIS and the EU Marie Curie
Fellowship.



\end{document}